\documentclass[manuscript,screen]{acmart}
\AtBeginDocument{%
  \providecommand\BibTeX{{%
    \normalfont B\kern-0.5em{\scshape i\kern-0.25em b}\kern-0.8em\TeX}}}

\setcopyright{acmcopyright}
\copyrightyear{2022}
\acmYear{2022}
\acmDOI{}

\acmConference[3rd KDD Workshop on Data-driven Humanitarian Mapping]{August 15, 2022}{Washington, DC}{USA}

\acmBooktitle{Proceedings of the 3rd KDD Workshop on Data-driven Humanitarian Mapping, August 15, 2022, Washington, DC USA} 

\begin{document}

\title{Multi-AI Complex Systems in Humanitarian Response}

\author{Joseph Aylett-Bullock}
\authornote{joseph@unglobalpulse.org}
\orcid{0000-0001-7551-3423}
\affiliation{%
  \institution{United Nations Global Pulse}
  \streetaddress{}
  \city{New York}
  \state{New York}
  \country{USA}
  \postcode{10017}
}

\author{Miguel Luengo-Oroz}
\email{miguel@unglobalpulse.org}
\orcid{0000-0002-8694-2001}
\affiliation{%
  \institution{United Nations Global Pulse}
  \streetaddress{}
  \city{New York}
  \state{New York}
  \country{USA}
  \postcode{10017}
}

\renewcommand{\shortauthors}{Aylett-Bullock}

\begin{abstract}
AI is being increasingly used to aid response efforts to humanitarian emergencies at multiple levels of decision-making. Such AI systems are generally understood to be stand-alone tools for decision support, with ethical assessments, guidelines and frameworks applied to them through this lens. However, as the prevalence of AI increases in this domain, such systems will begin to encounter each other through information flow networks created by interacting decision-making entities, leading to multi-AI complex systems which are often ill understood. In this paper we describe how these multi-AI systems can arise, even in relatively simple real-world humanitarian response scenarios, and lead to potentially emergent and erratic erroneous behavior. We discuss how we can better work towards more trustworthy multi-AI systems by exploring some of the associated challenges and opportunities, and how we can design better mechanisms to understand and assess such systems. This paper is designed to be a first exposition on this topic in the field of humanitarian response, raising awareness, exploring the possible landscape of this domain, and providing a starting point for future work within the wider community.
\end{abstract}




\maketitle

\section{Introduction}

Humanitarian emergencies are growing in number and scale, and in many cases interact with each other, adding to an increased level of complexity. AI decision support systems and automated decision-making technologies are increasingly being used in the humanitarian sector to address the many pressing challenges of responding to emergency situations, including: early warning and preparedness systems; assessment and monitoring capacities; service delivery and support; and operational and organizational efficiency \cite{spencer_humanitarian_2021}. Many of these efforts are in their nascent stages, however, there is a growing number that are becoming operational and acting as decision support, or decision-making systems in real-world situations.

Deployed AI systems are used at multiple levels of humanitarian decision-making, from headquarters (HQ) to field operations. However, they are typically considered to act in isolation. For example, HQ organizations may provide maps generated using AI systems \cite{logar_pulsesatellite_2020, nemni_fully_2020}, response teams may use Natural Language Processing (NLP) tools to collect feedback from affected populations to inform operational responses \cite{gill_experiments_2021}, and near and long term future scenarios may be forecasted using AI systems for resource mobilization and contingency planning \cite{martini_nowcasting_2021, foini_on_2021}. (For more examples see the following survey \cite{un_activities_2021}.) There is also a large movement focused on forecast-based financing which uses forecasting methods to inform funding of crisis prevention, mitigation, and response efforts \cite{forecast_financing}.

With a growing number of entities using AI systems to inform their response efforts, it is natural that these systems begin interacting with each other, despite the fact that  inter-dependencies are not always recognised, traceable, or addressed by design. In addition, these systems are often not isolated from the broader environment of AI systems beyond the humanitarian space. There are a multitude of other AI systems which externally impact the work of humanitarian organizations - e.g. social media algorithms which curate content and may effect analyses of social media trends and sentiments, or maps produced by external entities \cite{sirko_continental-scale_2021, ai_microsoft}.

There are numerous works highlighting the risks and potential harms associated with using AI systems in the humanitarian sector. These risks and harms include: questions of biases and fairness which can oppress minority voices; accountability of systems and the lack of legal frameworks; challenges surrounding transparency and human oversight; a lack of expertise in the proper and appropriate deployment and use of AI; data privacy and protection concerns; and the meaningful participation of affected communities in every stage of the AI lifecycle, as well as their ownership over such systems \cite{spencer_humanitarian_2021, coppi_explicability_2021, ocha_digital_2021, toplic_ai_2020}. In an attempt to address some of these concerns, there has been a drive to draft AI ethical guidelines, frameworks, and training materials at the international \cite{unesco_draft_2021},  national, and sector specific levels \cite{jobin_global_2019}.  These principles generally approach the problem of the ethical use of ‘AI systems’ in isolation, rather than from a systems perspective - i.e. again assuming such AI systems interact with their environment, but not explicitly considering an environment which may include other AI systems.

The field of multi-agent reinforcement learning (RL) has taken important steps in understanding the technical and theoretical behaviour of certain multi-AI systems \footnote{Indeed, multi-AI systems such as those discussed in this paper are occasionally referred to as `multi-agent' systems in which an `agent' can have the form of any general AI system. `Agents' in some literature can also refer to humans or AI systems interchangeably. In this paper, we use the phrase `multi-AI' for the avoidance of doubt as we are referring to AI systems in general, which may or may not have RL components and may have humans-in-the-loop. In the multi-agent RL literature, agents can also sometimes be considered to be of the same `class' - i.e. operating under the same or variations of the same set of rules and objectives - whereas in our setting the different actors/models/systems can be very heterogeneous in nature. Further, we also include humans in the information flow networks and feedback cycles referred to in this paper; however, we make explicit reference to them when we do so.}. Studies commonly focus on the deployment of multiple reinforcement learning models in a closed environment (`games') to monitor their interactions as they perform certain tasks, optimising for various objectives. Traditionally, RL agents have competed against each other in such games, with recent advances in setups where performance in the competition can be clearly defined (e.g. in games of skill), as well as in the more challenging context of zero-sum games \cite{yang_overview_2021, balduzzi_open_2019}. There are also many examples in which agents are not intended to be competing. Cooperative AI is an important consideration in areas such as self-driving cars, in which AI agents should work together and any negative competition could be dangerous (see \cite{dafoe_open_2020} for a range of open questions, challenges and opportunities around building cooperative AI systems).



In the real world, however, outside these closed game setups, there are erratic behaviours and social dilemmas at multiple scales which AI models interacting with these environments must accommodate. Methods, such as that proposed by Baker \cite{baker_emergent_2020}, have begun to try to address this challenge through the training of agents with randomised uncertain social preferences, updating the rewards of RL agents with uncertainty to mimic complex mixed-motive environments. Further work is still needed to take this into a real-world setting for stress testing.

The example of multi-agent RL is a simplified version of reality for simulating and understanding different behavioural patterns. In this paper, we discuss the broader context of AI systems which may or may not take the form of fully-automated RL agents, and in which humans also play important roles in decision-making chains and interactions. In the field of humanitarian response, there are a large number of actors, with mixed, numerous, and often hard to quantify objectives for which to optimise, working at multiple scales. These systems and actors may work more slowly than traditional RL agents in a sandbox environment, however, still react very rapidly by human decision-making standards. 

As we continue to operationally deploy AI systems (which are themselves complex systems) in complex humanitarian environments (which contain other AI systems), such systems become intertwined and interact, creating second-order effects which should be explored and understood. This mirrors the higher-order complexity seen when multiple interacting crises occur - e.g. when a conflict, food crisis and pandemic all exacerbate each other. These multi-AI complex systems can generate new sets of emergent behavior which will not be encompassed by only considering AI systems individually, and therefore requires new considerations vis-a-vis their impacts, risks and harms, ethics and overall assessments. There is much we can learn from the technical literature in the multi-agent RL and other spaces, and multidisciplinary approaches are necessary, and need to be developed quickly, to better understand this rapidly evolving multi-AI humanitarian landscape.


\section{Example of AI systems embedded into a humanitarian response pipeline}

A key starting point for understanding multi-AI systems and ensuring their behavior is understood and controlled is to piece interactions apart by considering such systems as information networks where decision-making nodes can be human, machine, or a combination of both.

\begin{figure}
    \centering
    \includegraphics[width=0.69\textwidth]{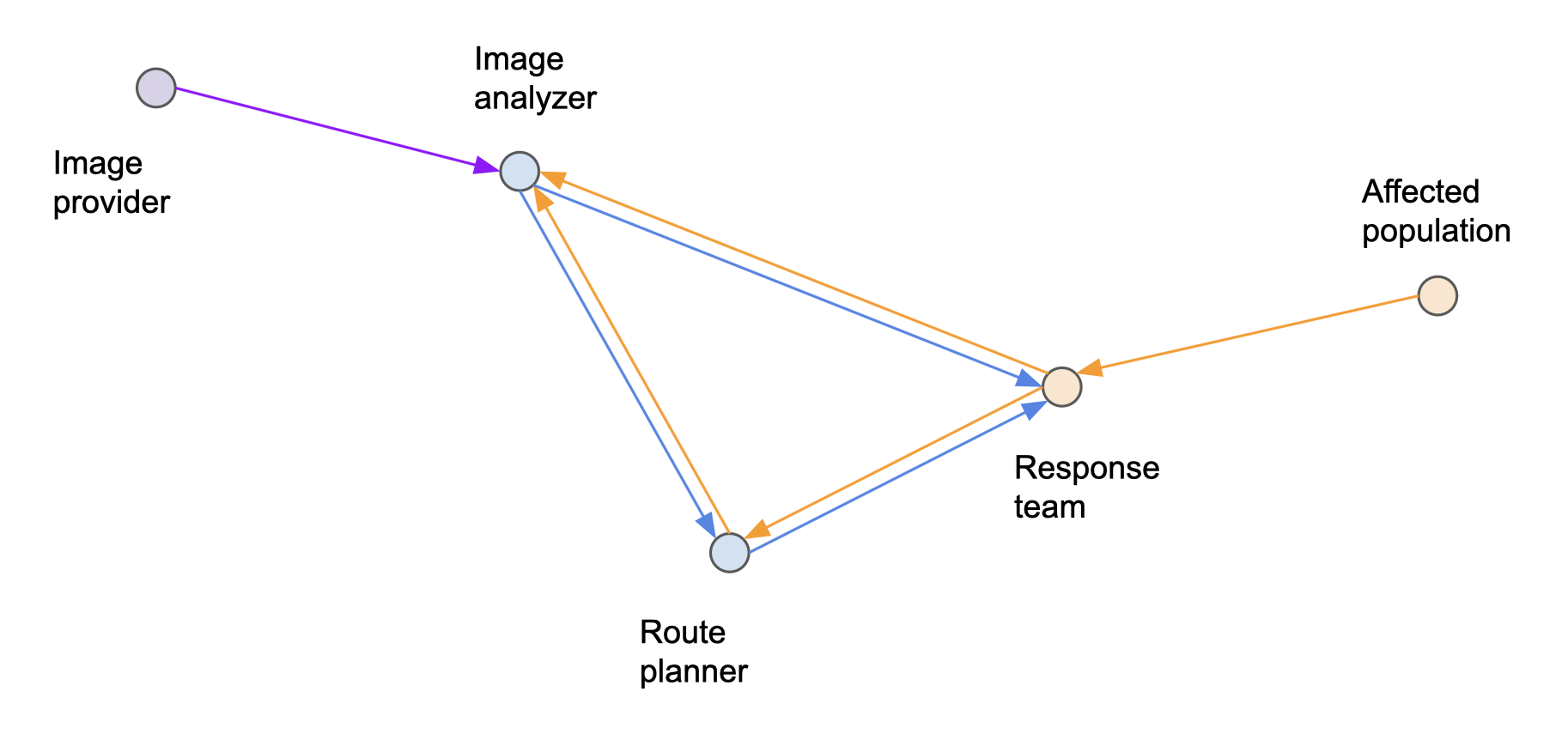}
   \caption{Example of a simple information flow network in a humanitarian response. Purple nodes represent AI-only systems, orange nodes represent human-only decision-makers, while blue nodes represent entities or groups which consist of both humans and AI systems together. The arrows represent information flow following the same color scheme.}
   \label{fig:network}
\end{figure}

It is instructive to start with an example of what such a system might look like given current real-world workflows. In Figure \ref{fig:network} we see an example of a simple information flow network in a realistic humanitarian response scenario. When a crisis occurs, such as a flood which affects a large population, it is common to use satellite imagery to assess the extent of the flood and/or damage. The image is usually provided by an external provider - through fully automated interactions - some of whom use AI models to process their imagery (e.g. to select those with minimal cloud cover or correct for cloud cover). This is passed to a human image analyzer who uses AI to help rapidly detect where the flood has occurred. The resulting flood map is then passed to another team who uses it as an input to an AI model for identifying the most affected population, and plan a route to their location. The plan is passed to a response team in order to deliver the emergency aid, and we assume that they also get the underlying flood map which is fed into the route planning model from the image analyzer. 

At various stages, feedback may be given based on the results. The affected population might give feedback to the response team telling them they have gone to the wrong place or that there are better ways to get to a given area. The response team might pass this information back to the route planners; however, if they consider it an error in the flood map, which they also have access to, they might give feedback directly to the image analyzer who produced it. The route planning team might also see errors in the flood map, based on the feedback from the response team which may or may not have also been provided directly to the image analyzer, and provide this feedback back to the image analyzer themselves. We regard it unlikely in this scenario that the image analyzer would provide feedback directly to the external image provider.

It is important to understand both kinds of information flows: the decision/output and the feedback. AI derived information can be erroneous and can cause the multitude of challenges, risks and associated harms generally elicited through existing AI system risk assessments and enumeration in ethical frameworks. When such erroneous information flows into other systems which then make use of it, such risks and harms can be compounded.

The feedback cycles can also be erroneous and can perpetuate or induce later incorrect results. While the process of model validation and updating based on feedback can be rigorous and systematic at the level of the individual AI system, the process of how, and to whom, an entity provides feedback becomes complex when there are multiple interacting systems. When working within complex systems, it can be hard for those giving feedback to correctly identify the system that generated the data \footnote{similar to the credit assignment problem in the RL literature}, and if a correction or validation is performed on the wrong system, this can further introduce errors through the feedback process itself.

In the example discussed above, the response team may incorrectly deduce that the flood map has errors, rather than the output of the route planning model, based on their own experience or feedback from the population, and therefore initiate a correction to the image analyzer’s model which results in feeding incorrect information back into the analyzer’s model rather than the route planner's model. This in turn leads to the image analysis model performance decreasing, which affects the route planning model input and in turn provides the response team with worse information. Taking this one step further, the response team may then find another error but this time report it to the route planning team when the error was actually due to the flood map. 

The above scenario is relatively simple, and humanitarian response workflows in reality can be much more complex with large numbers of actors and only partial feedback cycles. Nonetheless, by playing out this simple scenario, we can see that while each stage of the process may be correct in isolation - for example, each model may have passed ethical risk assessments and have validation and update procedures in place based on feedback - the broader multi-AI system may descend into a state of emergent erratic behavior with models becoming tuned to incorrect data due to an ill-understood landscape of information flow and feedback cycles. Note that this scenario can play out even under the assumption that each individual node which is using an AI system believes their input data to be reliable unless observed directly otherwise. Indeed, such behavior is not the only challenge to handle and understand, more of which will be discussed below.

\section{Towards trustworthy multi-AI systems: challenges and opportunities}

These multi-AI systems can present a range of new challenges, risks and associated harms, as well as opportunities for better understanding complex humanitarian situations more holistically, rather than compartmentally. From the discussion above, it is clear that one risk is that such systems become trained on erroneous data - e.g. with errors from one AI model feeding into another, or human errors in which system is updated and how (incorrect feedback pathways) - but which is thought to be reliable and can result in providing incorrect information to decision-makers during emergency situations and have significant harmful repercussions. Further, these complex interactions also have the potential to significantly enhance the effects of outliers. It is commonly known that AI systems can perform erratically on input data which is in the region of low or no statistics in the training domain; however, the training domain for the system as a whole can now be considered as a complex distribution over many interacting training domains \footnote{The complexity of these distributions not only lies in the number of individual interacting training domains, but in the way in which they may be statistically combined - e.g. due to the output of certain models feeding into others together with the effect of feedback loops. This is a broad and complex topic which is beyond the scope of this paper to investigate but which we see great value in further exploration.}. This can have negative impacts on the inherent uncertainties of the multi-AI system and therefore, if not understood and communicated properly, can result in ill-informed/over-confident decisions being made.

More broadly, by not understanding the complex interactions of these systems, efforts to include community participation in the design and deployment of AI systems becomes increasingly difficult. Indeed, the ability to ensure affected populations have meaningful ownership of AI systems presents further challenges as inputs to any end-user focused system may be unknowingly altered through these interactions. 

As mentioned above, multi-AI systems present key challenges to existing AI ethical principles, frameworks and guidelines. Decisions lack traceability unless the one/two-way interactions between systems, and feedback loops with participants, are understood, and this can have repercussions with regards to accountability and responsibility of decision-making, as well as the ability to have oversight over AI systems. For example, in the scenario laid out above, while the route planning team may be theoretically accountable for the decision of which route to direct the response team along, when responding rapidly in crisis situations, the route planning team may be able to legitimately claim that they were acting on information which was correct to the best of their knowledge as the full multi-AI system had not been understood.

Finally, while there are significant challenges, risks and harms which can be associated with such multi-AI systems, especially when they are ill-understood, there is still great potential for their use. Humanitarian situations, along with many other real-world scenarios in which AI systems can and will be used, are growing in their complexity and such multi-AI systems may be able to help explain and interact with a broader collection of crises to understand them more holistically.

\section{Discussion and next steps}

Multi-AI systems can be thought of as a complex interacting network of interconnected complex systems. These systems are being increasingly used in many domains, including in humanitarian response, but their behaviours are often not understood in their entirety. However, if safely and successfully deployed, multi-AI systems have the ability to help answer new questions, better understand much of the increasing complexity of our world and provide new solutions to operational challenges. 

To understand these higher-order complex systems we need a shift in the modalities of thinking and operating as a community, which includes understanding a broad and evolving spectrum of system interactions at different scales. We need to expand the traditional assumptions of data generation and transformation, considering how information is shared and where it is coming from, as well as how feedback cycles are created, either formally or informally. In settings such as humanitarian emergencies there are many sets of actors linked through multiple pathways, rather than acting as linear systems, which need to be understood in order for them to function effectively. To address these challenges, we propose several areas of future research.

First, from a methodological perspective, a continuation of the development of frameworks for building and testing multi-AI systems, as well as for understanding emergent behaviours of such systems is essential. Working closely with researchers in other fields, such as the multi-agent RL community, will be essential to bringing these theoretical works into critical real-world domains.

When embarking on projects, mapping the network of information flows to understand the broader multi-AI system environment is key. In the case of multi-AI systems in humanitarian domains, network `owners' or `regulators' could be assigned who have the responsibility to oversee the system (even if they do not `own' or `control' each individual component). The development of stress tests for these systems as a whole, involving all the entities in the network who are providing inputs to relevant subsystems, can then help understand possible erroneous and emergent behavior which can be mitigated before harm is caused. This can be made easier through clear communication, by AI model-creators and owners, of data inputs and model versioning for systems used. This includes the data inputs received by other AI systems and, at each decision node, one should be able to validate any AI augmentation to the input data which has occurred.

Secondly, a greater degree of cross-pollination of these technical communities with legal, ethics, and humanitarian response experts is needed to create ethical and human rights frameworks appropriate for multi-AI systems. A systematic mapping exercise to understand the potential gaps in existing ethical principles and guidelines in the context of multi-AI systems can be supported with technical exercises to both quantify such principles and guidelines and measure when certain criteria are met at the individual AI system level, and then fail to be met at the multi-AI system scale - i.e. questions such as: `Under which conditions a multi-AI system is unsafe even if the individual subsystems are safe?' and `Is a multi-AI system interpretable when every AI subsystem is interpretable?'. The multi-agent systems and AI communities have been exploring some of these concepts, and continuing to broaden the scope of this work into the field of AI safety, fairness and others is essential for ensuring both a theoretical and practical understanding of such behaviours.

Similarly, risk and harms assessments which are currently used to evaluate and mitigate risks of AI systems in humanitarian contexts - both before and during project implementations - should be adapted to work with different scales of complexity of the multi-AI systems. One could draw a parallel to how risks and harms assessments have been adapted to include not only individual privacy assessments but also group privacy assessments - as protecting individual privacy does not necessarily mean protecting a particular ethnic group or vulnerable population \cite{UNGP_risks_harms_1, UNGP_risks_harms_2}.

Third, from an operational and practical perspective, methods for communicating these concepts to project managers and decision-makers in the humanitarian field are needed to socialise the challenges and risks associated with multi-AI systems. Mapping detailed examples of multi-AI systems is also fundamentally needed. While we have presented a specific case study of a multi-AI system in a humanitarian context, there are many other such examples in humanitarian response. For example, aid based on cash transfer mechanisms might use, in addition to multiple human steps, a number of population estimation methods (some based on indirect measures provided by AI models), AI models for disaster impact assessments, and biometric systems for beneficiary identification. Alternatively, a multi-AI system for emergency communications around mis- and dis-information could be composed of human analysts, communication officers and AI subsystems for data collection, language models with biases for certain languages, and AI-driven alert detectors. Detailed examples will help build operational guidelines, as well as capacity building exercises, and communities of practice could also support the sharing of knowledge and information around this topic within the context of emergency response.

Finally, more broadly, as a multidisciplinary community, greater collaboration is needed to understand the complexities and associated risks and harms of multi-AI systems, both within the humanitarian domain and beyond. There is a wealth of other fields and literature to be drawn on - including: complex systems, multi-agent modelling, systems engineering, and reinforcement learning - which can be adapted to humanitarian contexts.

As AI continues to be rapidly and widely deployed in the humanitarian domain and beyond, these present and emerging challenges will continue to grow and develop. It is therefore vital that we act before multi-AI systems, many of which are already deployed without proper design and testing, become sufficiently embedded that harm is created and future harms become challenging to prevent.








\section{Disclaimer}
The authors alone are responsible for the views expressed in this article and they do not necessarily represent the views, decisions or policies of the institutions with which they are affiliated including the United Nations.

\section{Acknowledgments}
United Nations Global Pulse work is supported by the Governments of Sweden and Canada, and the William and Flora Hewlett Foundation. The authors are grateful to Katherine Hoffmann Pham and Akbir Khan for their useful discussions and comments while preparing this work.

\bibliographystyle{ACM-Reference-Format}
\bibliography{bib}

\end{document}